\documentclass[12pt]{article}
\usepackage{rotate}
\usepackage{epsfig}
\usepackage{psfrag}
\usepackage{a4}

\begin{document}

\newcommand \be  {\begin{equation}}
\newcommand \bea {\begin{eqnarray} \nonumber }
\newcommand \ee  {\end{equation}}
\newcommand \eea {\end{eqnarray}}

\title{\bf Statistical properties of stock order books: empirical results and models}
\author{Jean-Philippe Bouchaud$^{\dagger,*}$, Marc M\'ezard$^{**,*}$, Marc Potters$^*$}
\maketitle
{\small
{$^\dagger$ Commissariat \`a l'Energie Atomique, Orme des Merisiers}\\
{91191 Gif-sur-Yvette {\sc cedex}, France}\\

{$^*$ Science \& Finance, CFM, 109-111 rue Victor Hugo}\\
{92 353 Levallois {\sc cedex}, France}\\

{$^{**}$ Laboratoire de Physique Th\'eorique et Mod\`eles Statistiques}\\
{Universit\'e Paris Sud, Bat. 100, 91 405 Orsay {\sc cedex}, France}\\
\date{\today}
}

\begin{abstract}
We investigate several statistical properties of the order book of three liquid 
stocks of the Paris Bourse. The results are to a large degree independent of the stock studied. 
The most interesting features 
concern (i) the statistics of incoming limit order prices, which follows a power-law 
around the current price with a diverging mean; and (ii) the 
shape of the average order book, which can be quantitatively reproduced using a `zero 
intelligence' numerical model, and qualitatively predicted using a simple approximation. 
\end{abstract}

Financial markets offer an amazing source of detailed
data on the collective behaviour of interacting
agents. It is possible to find many reproducible patterns and 
even to perform experiments, which bring this atypical subject into the realm 
of experimental science. The situation is simple and well defined, 
since many agents, with all the same goal, trade the very same asset. 
As such, the statistical
analysis of financial markets also offers an interesting testing 
ground not only for economic theories, but also for more ambitious 
theories of human activities. 
One may indeed wonder to what extent it is necessary to invoke human
intelligence or {\it rationality} to explain the various universal 
statistical laws which have been recently unveiled by the systematic 
analysis of very large data sets.

Many statistical properties of financial markets have already been explored, 
and have
revealed striking similarities between very different markets (different traded assets,
different geographical zones, different epochs) \cite{Cont,MS,BP}. 
For example, the distribution of price changes
exhibits a power-law tail with an apparently universal exponent \cite{Gopi1,Lux,Dacorogna2,
Longin}. 
The volatility of most assets
shows random variations in time, with a correlation function which decays as a small power of 
the time lag, again quite universal across different markets \cite{BMD}. 

Here, we study the statistics
of the order book, which is the ultimate `microscopic' 
level of description of financial
markets. 
The order book is the list of all buy and sell `limit' orders, with their 
corresponding price
and volume, at a given instant of time. 
A limit order specifies the maximum (resp. minimum) 
price
at which an investor is willing to buy (resp. sell) a certain number of shares (volume). At 
a given instant of time, all limit buy orders are below the best buy 
order called the {\it bid price}, while all sell orders are above the best 
sell order called
the {\it ask price}. When a new order appears (say a buy order), 
it either adds to the book 
if it
is below the ask price, or generates a trade at the ask if it is above 
(or equal to) the ask 
price (we call all these `market orders' even if technically they could
also be `marketable limit orders').
The price dynamics is therefore the result of the interplay 
between the order book and the order flow. The study of the order book is very 
interesting both for academic and practical reasons. It provides intimate 
information on the processes of trading and price formation, and reveals a 
highly non trivial structure of the agents' expectations (see below): as such, it is 
of importance to test some basic notions of economics and models of market
microstructure. The practical motivations are also clear. Regulatory bodies want to set the
rules of exchange markets such as to produce fair, orderly trading that maximize flow. For market participants, issues such as the market impact or the relative merit of 
limit versus market orders are determined by the structure and dynamics of the
order book.

Complete data on the order book of certain markets, such as the Paris 
Bourse (now Euronext), 
is now available,
and contains a particularly abundant information. {\it All} 
orders (representing 
hundreds of thousands orders per month on liquid stocks) on 
{\it all} stocks are 
stored, which
makes possible the reconstitution of the full order book at any instant 
of time, including orders that are not directly observable on traders' screens.

Many questions can be studied; the systematic investigation
of these data sets is only very recent \cite{Biais,Maslov,Challet} and has motivated a number
of interesting theoretical work \cite{Bak,Kogan,Maslov,Aussie,Slanina,Farmer}. Here, 
we mostly focus on `static'
properties of the order book, such as the distribution of incoming limit orders, the 
average shape of the order book in the moving reference frame of the price, or the distribution
of volume at the bid/ask. Many other `dynamical' properties can also be analyzed, such as 
the response of the price to order flow, the full temporal correlation of the book, etc.
In order to keep our study reasonably focused, we leave these dynamical aspects, that we 
have not fully understood yet, for a subsequent publication. 

Our main results are as follows: (a) the price at which new limit orders are placed is,
somewhat surprisingly, very broadly (power-law) distributed around the current bid/ask; 
(b) the average order book has a maximum away from the current bid/ask, and a tail 
reflecting the
statistics of the incoming orders; (c) the distribution of volume at the bid (or ask)  
follows a Gamma distribution. Notably, we find exactly the same statistical features, 
for the three liquid stocks studied. We then study numerically
a `zero intelligence' model of order book which reproduces most of these empirical results. 
Finally, we show how the characteristic shape of the average order book can be analytically
predicted, using a simple approximation.

The data provided by Paris Bourse gives the history of all transactions, with their 
price, volume and time stamp, of all quotes (bid and ask prices) with the corresponding 
volumes, and of all orders, with their price, volume and time stamp. This information 
allows in principle to reconstruct the whole order book at any instant of time. However, some limit orders (roughly $10 \%$ of them) are 
modified or cancelled before being executed; unfortunately the data base only contains this qualitative information, but not the time at 
which a given order is modified/cancelled, nor the resulting new price. 
When reconstructing
the order book at a given instant of time, we have therefore made two extreme assumptions, 
which
lead to nearly identical conclusions. Either we discard these orders altogether, or we keep
them until the time where we can be sure that they have been previously modified/cancelled, 
otherwise they would
have been executed since the transaction price was observed to be below (for buy orders) or
above (for sell orders) the corresponding limit price. The results given below uses the
second procedure. We had available the data corresponding 
to all stocks during February 2001, from which we have extracted three of the most liquid
stocks: France-Telecom (F.T.), Vivendi and Total. Some basic information is summarized in Table I
for France-Telecom and Total. We expect that our findings will not depend on the
particular month considered, but that some differences may appear when one studies stocks 
with smaller capitalisation (for example the average bid-ask spread is much larger): these 
questions will be investigated in the near future.
Let us call $a(t)$ the ask price at time $t$ and $b(t)$ the level of bid price at time $t$. The midpoint $m(t)$ is the average between the bid and the ask: $m(t)=[a(t)+b(t)]/2$.
The smallest possible change of these quantities, called the `tick', was $0.05$ Euros for
France-Telecom and Vivendi, and $0.10$ Euros for Total. 

\begin{table}
\begin{center}
\begin{tabular}{||l|c|c||} \hline\hline
Quantity\ \hspace{0.5cm} \   & \hspace{0.5cm} F.-T. \hspace{0.5cm} &
\hspace{0.5cm}  Total \hspace{0.5cm} \\ \hline
Initial/final price (Euros)  &  90-65 &   157-157 \\ \hline
Tick size (Euros)  &  0.05 &  0.10 \\ \hline
Total \# orders  &  270,000 &  94,350  \\ \hline
\# market orders  &  28,600 &  9,300 \\ \hline
\# trades  &  176,000 &  60,000 \\ \hline
Transaction volume  &  75,6\, $10^6$ &  23,4\, $10^{6}$ \\ \hline
Average bid-ask (ticks)  &  2.0 &  1.4 \\ \hline
Average volume at bid/ask  &  2700 &  2400 \\ \hline\hline
\end{tabular}
\end{center}
\caption[]{\small Some useful data for the studied stocks 
(February 2001). The transaction volume is in number of shares. The figures for 
Vivendi are nearly identical to those for France-Telecom.
}
\end{table}

We denote by $b(t)-\Delta$ the
price of a new buy limit order, and $a(t)+\Delta$ the price of a new sell limit order. 
Notice that $\Delta$ can be negative 
(this is the only case where the spread $g(t)=a(t)-b(t)$ can be narrowed),
but is always larger than $-g(t)$ (otherwise it would be a market order).
A first interesting question concerns the distribution 
density of $\Delta$, i.e. the distance 
between the current price and the incoming limit order. We find that $P(\Delta)$ is identical
for buy and sell orders (up to statistical fluctuations); the shape of $P(\Delta)$ is
found to be very well fitted (see Fig. 1) by a power-law:
\begin{equation}
\label{PDelta}
P(\Delta) \propto \frac{\Delta_0^\mu}{(\Delta_1+\Delta)^{1+\mu}}, \qquad \Delta \geq 1
\end{equation}
with an exponent $\mu \approx 0.6$ for all three stocks. This power-law extends from $1$
tick to over $100$ ticks (sometimes even $1000$ ticks), corresponding to a relative 
change of price of $5 \%$ to $50 \%$ \cite{Farmerbis}.
There are also orders placed at the bid (or ask) or 
within the spread. One finds that $P(\Delta=-1) \sim
P(\Delta=0) \sim P(\Delta=1)$, and these orders add up to roughly half of the total number of
orders.
The distribution of orders Eq.~(\ref{PDelta}) has a maximum around the current price 
(this was already noted in \cite{Biais}). But the fact that 
$\mu < 1$ means that the average $\Delta$ is formally infinite (although of 
course the distribution is ultimately cut-off for large $\Delta$'s).

\begin{figure}
\begin{center}
\epsfig{file=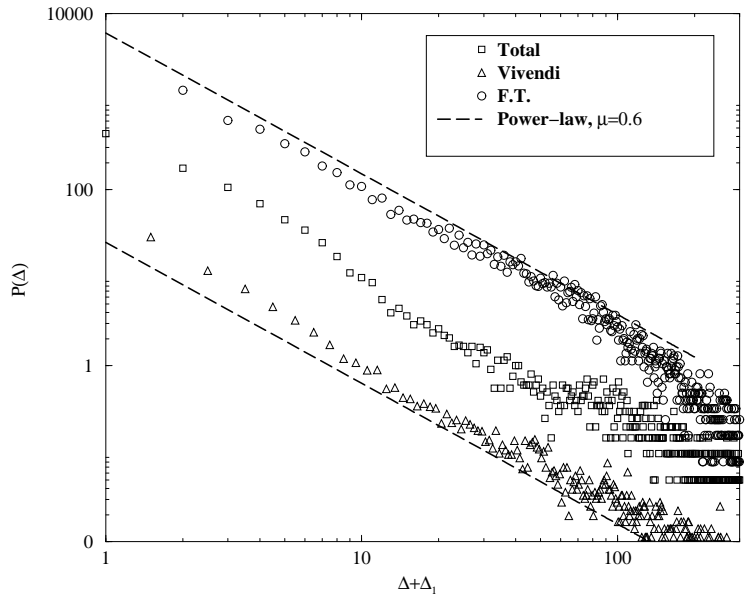,width=8cm,angle=270}
\end{center}
\caption{\small Number of incoming orders arriving at a distance $\Delta$ from the
current best price, as a function of $\Delta_1+\Delta$ (in ticks), in log-log coordinates. We chose $\Delta_1 = 1$ for F.T.,
$0.5$ for Vivendi and $0$ for Total. We also found $P(\Delta=-1) \sim
P(\Delta=0) \sim P(\Delta=1)$ (not shown). The symbols correspond to 
buy orders, but sell orders show an identical distribution. The straight lines correspond to
$\mu=0.6$. Note that the
power-law crosses over to a faster decay beyond $\Delta \approx 100$.  \label{fig1} }
\end{figure}

It is quite surprising to observe such a broad distribution of limit order prices, which
tells us that the opinion of market participants about the price of the stock in a 
near future could be anything from its present value to $50 \%$ above or below this
value, with all intermediate possibilities.
This 
means that market participants believe that large jumps in the price of stocks 
are always possible, and place orders very far from the current price in order to
take advantage of these large potential fluctuations. (Note that placing a limit 
order is free of charge.) A naive argument would then suggest that the
probability to place an order at distance $\Delta$ should be proportional to the
probability that the price moves more than $\Delta$ in order to meet the order. 
Since the tail of the distribution of price increments is a power-law with an exponent 
$\mu_{\delta p} \approx 3$ \cite{Gopi1}, this would 
indeed lead to a power-law for $P(\Delta)$, but with a value  
$\mu = \mu_{\delta p}-1 \approx 2$ larger than the observed one. This however does not take into
account the fact that market participants have different time horizons -- large $\Delta$'s 
presumably correspond to more patient investors. The value of $\mu$ should result from an
interplay between the {\it perceived} distribution of future price changes and the distribution
of time horizons, which could well be itself a power-law. We feel that more theoretical work is
needed to fully account for this striking result.

Limit orders strongly vary in volume. We find that the unconditional limit order size,
$\phi$,
is distributed uniformly in log-size, between $10$ and $50,000$ 
(both for buy or sell orders). One can study the correlation between the incoming {\it volume} $\phi$ (number of shares for a given order) and the distance $\Delta$ between the order 
and the current price. We find that the conditional
averaged volume $\langle \phi \rangle|_\Delta$ is roughly independent of $\Delta$ between
$1$ and $20$ ticks, but decays as a power law $\Delta^{-\nu}$, with $\nu \sim 1.5$  
beyond $\Delta^*$ (with $\Delta^* \approx 20$ ticks for FT, and $\approx 50$ ticks for
the other two stocks). Not
unexpectedly, extremely far limit orders tend to be of smaller volumes. Since the
average volume arriving at $\Delta$ is equal to the number of orders at $\Delta$ 
multiplied by $\langle \phi \rangle|_\Delta$, one predicts, using the shape of $P(\Delta)$,
that the distribution of incoming volume should then decay as $\Delta^{-1-\mu}$ 
for $\Delta < \Delta^*$ and as $\Delta^{-1-\mu-\nu}$ for large $\Delta > \Delta^*$, a
feature that we have confirmed directly.

We now turn to the shape of the order book. The order flow is maximum around the
current price, but 
an order very close to the current price has a larger probability of being executed and 
thus disappear from the book. It is thus not a priori clear what will be the 
average shape of the  order book. Quite interestingly, we find that the (time-averaged)
volume of the queue in the order book (that we will call for simplicity the `average order book') is symmetrical, and has a maximum away 
from the current bid (ask): see Fig. 2. 
This was also noted in \cite{Biais}. This shape furthermore appears to be {\it universal},
 up to a rescaling 
of  both the $\Delta$ 
axis and the volume axis, at least for the three stocks studied. In order to test the universality of this result, we are currently studying less
liquid stocks, and in a different time interval. The empirical 
determination of the average order book is the central 
result of this
paper, and simple models that explain this shape will be discussed below.

\begin{figure}
\begin{center}
\epsfig{file=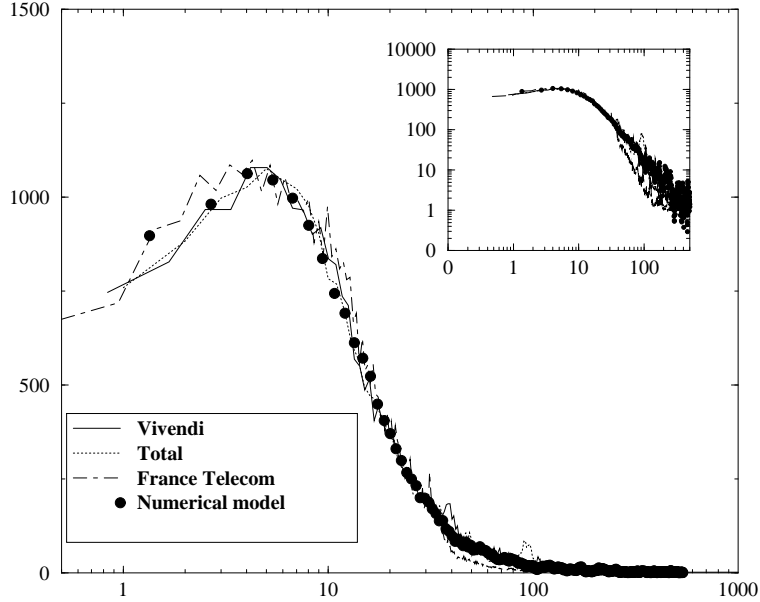,width=8cm,angle=270}
\end{center}
\caption{\small Average volume of the queue in the order book for the three stocks, as a function of the distance $\Delta$ from the current bid (or ask)
in a log-linear scale. Both axis have been rescaled in order to collapse the curves corresponding to the three stocks. The thick dots correspond to the numerical
model explained below, with $\Gamma=10^{-3}$ and $p_m=0.25$. Inset: same data in 
log-log coordinates.\label{fig2}}
\end{figure}

The next question concerns the volume fluctuations around this average shape. We first 
study the
distribution $R(V)$ of volume at the bid (or ask). Again, the two are 
identical, and can 
be fitted by a Gamma distribution for the volume (Fig. 3):
\be\label{gamma}
R(V) \propto V^{\gamma-1} \exp\left(-\frac{V}{V_0}\right).
\ee
We find $\gamma \simeq 0.7 - 0.8$ for all three stocks. A Gamma distribution with 
$\gamma \leq 1$ has its maximum for $V=0$. This shows that the {\it most probable value} of
the volume at bid is very small, although its typical value is quite large 
($V_0 \simeq 2700$ for France-Telecom).

\begin{figure}\begin{center}
\epsfig{file=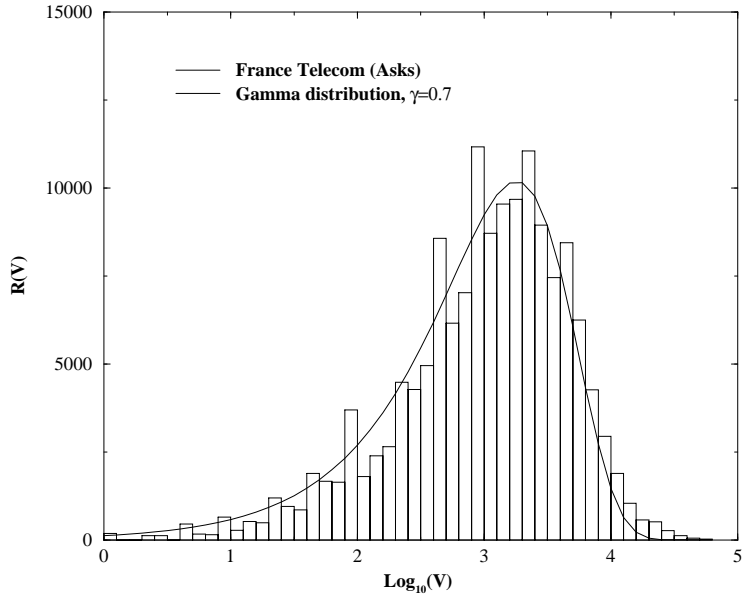,width=8cm,angle=270}
\end{center}
\caption{\small Histogram of the log-volume at the ask (same for bids), 
for France Telecom. The fit corresponds to Gamma distribution, Eq.~(\protect\ref{gamma}) (after a change of variables to $\log V$), with 
$\gamma=0.7$. \label{fig3} }
\end{figure}

Therefore, the order book has strong fluctuations and at a given instant of time can look 
rather different from its average, Fig. 2. We have also studied the fluctuations
of volume in the book as a function of $\Delta$, defined as:
\be
\sigma_V(\Delta)= {\sqrt{\langle V^2 \rangle|_\Delta - \langle V \rangle|_\Delta^2}}.
\ee
We find that $\sigma_V(\Delta)$ is of order $1$ for $\Delta=1$ (as expected from the 
above Gamma distribution), and is roughly constant up to $\Delta=50$. 
This shows that far from the current price, 
the instantaneous order book shows large {\it relative} fluctuations, which is indeed 
reasonable. We have also studied the full covariance matrix of the fluctuations 
$C_V(\Delta,\Delta')$ (note that $C_V(\Delta,\Delta) \equiv \sigma^2_V(\Delta)$). For France-Telecom, its first eigenvalue corresponds to 
a dilation of the book, whereas the second one reflects an even/odd oscillation (orders
are more numerous at prices in tenth of Euros than in twentieth of Euros). Higher 
eigenvalues are structureless, suggesting that the orderbook has no `elasticity'. This
is expected, since most investors do not have access to the order book in real time.

We now turn to a quantitative interpretation of the data. 
First, we have simulated an
artificial `zero intelligence' order book, much in the same spirit as in 
\cite{Challet,Farmer}, 
but more directly 
inspired from our empirical results. Limit orders, with a size distributed uniformly
on a log scale, are launched at random 
with an a priori `sprinkling' distribution $P(\Delta)$, which we take identical
to the empirical  one.  
Here we depart from \cite{Farmer}, where all orders have unit size and the 
`sprinkling' distribution is uniform between $0$ and a certain $\Delta_{\max{}}$. The
size distribution was actually found to play a minor role except in the far 
tail of the average order book, where the exponent $\nu$ introduced above becomes 
important. For example, we also studied the case of orders of equal size
with very similar results. 

On the other hand, choosing the 
correct `rain' distribution 
$P(\Delta)$ is crucial to reproduce the shape of the order book. We also launch
a certain fraction $p_m$ of market orders (chosen to be typically 
$1/6-1/4$th of the total) in order to
trigger trades. The results we find (in particular the averge size of the bid-ask spread) are quite sensitive to the value of $p_m$. Finally, with a constant probability $\Gamma$ (typically
of order $10^{-3}$) per unit time, 
independently of both size and position in the book, an order is cancelled. We have found that
this simple model is able to reproduce quantitatively 
many of the features observed in the empirical data,
such a the shape of the order book (see Fig. 2), or the Gamma distribution of volume at the bid/ask $R(V)$, with a similar value for the exponent $\gamma$. This
model also reproduces other empirical properties, such as the short time dynamics of 
the `midpoint' and the sublinear volume dependence of the response to an 
incoming market order \cite{Gopi,Farmer,ustocome}. However, we note that the above simple model overestimates the average size of the queue close to $\Delta=0$. This might be due either to 
the fact that our procedure to treat modified orders leads to an systematic bias which is large
near the best price, or that orders close to the best price have a non negligible strategic 
content that our model fails to capture. 

Finally, we discuss a simple analytical approximation which allows us to compute the average
order book from the ingredients of the numerical model. This was also attempted in 
\cite{Farmer}. Although very few details were given in \cite{Farmer}, our method appears to 
be quite different from theirs. 
Consider sell orders. Those at distance $\Delta$  from the current ask at time $t$ are 
those which were 
placed there at a time $t' < t$, and have survived until time $t$, that is, (i) have not 
been cancelled; (ii) have not been touched by the price at any intermediate time $t''$ 
between $t'$ and $t$. An order at distance $\Delta$ at time $t$ in the reference frame of
the ask $a(t)$ appeared in the order book at time $t'$ at a distance $\Delta+a(t)-a(t')$. 
The average order book can thus be written, in the long time limit, as:
\be
\rho(\Delta,t) = \int_{-\infty}^t {\rm d} t' \int du P\left(\Delta+u\right)
 {\cal P}
\left(u | {\cal C}(t,t')\right) 
{\rm e}^{-\Gamma(t-t')},
\ee
where  ${\cal P}\left(u | {\cal C}(t,t')\right)$
is the conditional probability that the time evolution of the price produces
a given value of the ask difference $u=a(t)-a(t')$, given 
the condition  that the path always 
satisfies $\Delta+a(t)-a(t'') \geq 0$ at all intermediate times $ t'' \in [t',t]$. 
The evaluation of ${\cal P}$ requires the knowledge of the statistics of the price process.
Because of the exponential cut-off ${\rm e}^{-\Gamma(t-t')}$, only 
the short time behaviour of the
process is relevant where the confinement effects of the order book on the price are 
particularly
important \cite{Maslov,Farmer,ustocome}. Nevertheless, to make progress, 
we will assume that the
process is purely diffusive. In this case, ${\cal P}$ can be calculated using the method of 
images. One finds:
\be
{\cal P}\left(u | {\cal C}(t,t')\right)=\frac{1}{\sqrt{2\pi D \tau}}
\left[\exp\left(-\frac{u^2}{2D\tau}\right)
-\exp\left(-\frac{(2 \Delta +u)^2}{2D\tau}\right) \right],
\ee
where $\tau = t-t'$ and $D$ is the diffusion constant of the price process.  

After a
simple computation, one finally finds, up to a multiplicative constant which only 
affects the overall normalisation of $\rho_{\rm{st}}(\Delta)=\rho(\Delta,t\to \infty)$:
\be
\rho_{\rm{st}}(\Delta)= {\rm e}^{-\alpha \Delta} \int_0^\Delta {\rm d} u P(u) \sinh(\alpha u)
+ \sinh(\alpha \Delta) \int_\Delta^\infty {\rm d} u P(u) {\rm e}^{-\alpha u},
\ee
where $\alpha^{-1}=\sqrt{D/2\Gamma}$ measures the typical variation of price during the
lifetime of an order, and fixes the scale over which the order book varies. When the 
distribution of order flows $P(\Delta)$ has the power-law shape Eq.~(\ref{PDelta}) with 
$\mu < 1$, the parameter $\alpha$ can be rescaled away in the `continuous' limit where $\alpha^{-1}$
is much larger than the tick size (which is the relevant limit for stocks, where 
$\alpha^{-1} \sim 10$). In this case,
the shape of the average order book only depends on $\mu$ and
$\hat \Delta=\alpha \Delta$, and  is given by the following 
convergent integral:
\be\label{final}
\rho_{\rm{st}}(\hat \Delta)={\rm e}^{-\hat\Delta} \int_0^{\hat\Delta} {\rm d} u \,u^{-1-\mu} \sinh(u)
+ \sinh(\hat\Delta) \int_{\hat\Delta}^\infty {\rm d} u \, u^{-1-\mu} {\rm e}^{-u}.
\ee
\begin{figure}
\begin{center}
\epsfig{file=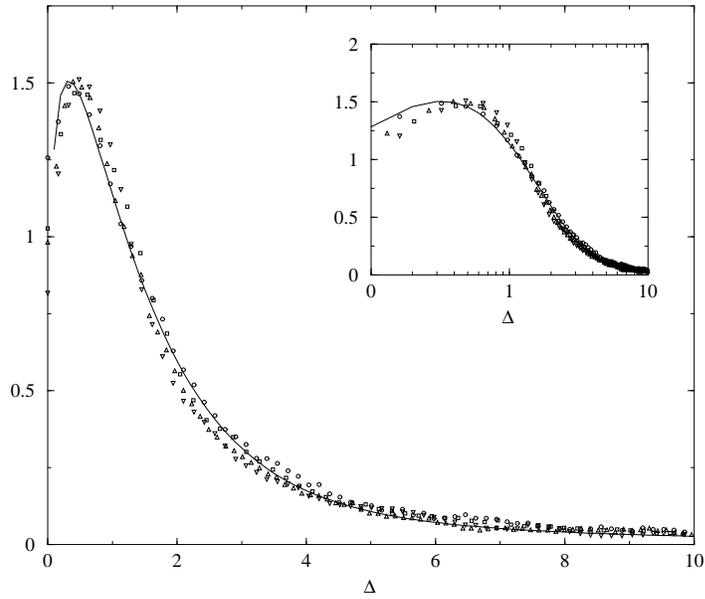,width=8cm,angle=270}
\end{center}
\caption{\small The average order book of the numerical model
with various choices of parameters ($\mu=.6$, $p_m\in \{ 1/4,1/6 \}$,
and $\Gamma \in \{10^{-3},5\, 10^{-4} \}$ is
compared to the approximate analytical prediction, 
(full curve), Eq.~(\ref{final}). 
After rescaling the axes, the various results roughly scale on the same curve,
which is well reproduced by our simple analytic argument. The inset shows the same 
data in a log-linear representation.
\label{fig4} }
\end{figure}

For $\Delta \to 0$, the average order book vanishes in a singular way, as 
$\rho_{\rm{st}}(\Delta) \propto \Delta^{1-\mu}$, whereas for
$\Delta \to \infty$, the average order book reflects the incoming flow of orders:
$\rho_{\rm{st}}(\Delta) \propto \Delta^{-1-\mu}$. We have plotted in Fig.\ref{fig4}
 the average order
book obtained numerically from
the `zero intelligence' model (which is very close to empirical data, see Fig. 2) and compared
it with Eq.~(\ref{final}), with $\mu=0.6$
and various choices of parameters. After rescaling the two axes, the
various numerical models lead to very similar average
order books, and the analytic approximation appears
rather satisfactory, bearing in mind the roughness of our `diffusive'
approximation.

The shape of the average order book 
therefore reflects the competition between a power-law flow of limit orders with a
finite lifetime, and the price dynamics that removes the orders close to the current price.
These effects lead to a universal shape which will presumably hold for many different markets,
provided the lifetime of orders is sufficiently long compared to the typical time between 
trades, and the volatility on the scale of the orders lifetime is somewhat larger than the 
ticksize. \footnote{Preliminary results indeed show that the average order book in futures
markets has a similar hump away from the current midpoint.} On the other hand, the detailed
shape of the book in the immediate vicinity of the best price (i.e., around $\Delta=0$) is not
expected to be universal.

In conclusion, we have investigated several `static' properties of the order book. We
have found that the results appear to be independent of the stock studied. 
The most interesting features 
concern (i) the statistics of incoming limit order prices, which follows a power-law around 
the current price 
with a diverging mean -- suggesting, quite surprisingly, that market participants believe that very large 
variations of the price are possible within a rather short time horizon; and (ii) the 
shape of the average order book, which can be quantitatively reproduced using a `zero 
intelligence' numerical model, and qualitatively predicted using a simple approximation. 
One of the most interesting open problems is, in our view, to explain the clear 
power-law behaviour of the incoming orders and the difference between our results and those
of \cite{Farmerbis}. 
The dynamical properties of the order book, and the way the price reacts to order
flow, will be the subject of a further study \cite{ustocome}.

\vskip 1cm
\noindent{Acknowledgements:} We thank Jean-Pierre Aguilar, Jelle Boersma, 
Damien Challet, J. Doyne Farmer, Laurent Laloux,
Andrew Matacz, Philip Seager and Denis Ullmo for stimulating and useful discussions.


\end{document}